\documentclass{aa} 
\usepackage{times}
\usepackage{psfig}
\usepackage{epsfig}

\begin{document} 


\title{Long-term evolution of FU\,Orionis objects at infrared
    wavelengths\thanks{Based on observations with ISO, an ESA project
    with instruments funded by ESA member states (especially the PI
    countries France, Germany, the Netherlands and the United Kingdom)
    with participation of ISAS and NASA.}}
\author{P. \'Abrah\'am\inst{1} 
        \and 
        \'A. K\'osp\'al\inst{2}
        \and 
        Sz. Csizmadia\inst{1}
        \and 
        M. Kun\inst{1}
        \and 
        A. Mo\'or\inst{1}
        \and 
        T. Prusti\inst{3}
        } 
\offprints{P. \'Abrah\'am, email: abraham@konkoly.hu}
\institute{Konkoly Observatory of the Hungarian Academy of Sciences,
           P.O. Box 67, H-1525 Budapest, Hungary 
      \and Department of Astronomy, E\"otv\"os Lor\'and University,
           P.O. Box 32, H-1518 Budapest, Hungary
      \and ESTEC/SCI-SAF, Postbus 299, 2200 AG Noordwijk, The Netherlands}
\date{Received date; accepted date} 
\authorrunning{P. \'Abrah\'am et al.}  
\titlerunning{Long-term evolution of FU\,Orionis objects at infrared
              wavelengths}
\abstract{We investigate the brightness evolution of 7 FU\,Orionis
systems in the $1{-}100\,\mu$m wavelength range using data from the {\it
Infrared Space Observatory} (ISO). The ISO measurements were
supplemented with 2MASS and MSX observations performed in the same
years as the ISO mission (1995$-$98). The spectral energy distributions
(SEDs) based on these data points were compared with earlier ones
derived from the IRAS photometry as well as from ground-based
observations carried out around the epoch 1983.
In 3 cases (Z\,CMa, Parsamian\,21, V1331\,Cyg) no difference between
the two epochs was seen within the measurement
uncertainties. V1057\,Cyg, V1515\,Cyg and V1735\,Cyg have become
fainter at near-infrared wavelengths while V346\,Nor has become
slightly brighter. V1057\,Cyg exhibits a similar flux change also in
the mid-infrared. At $\lambda\,{\geq}\,60\,\mu$m most of the sources
remained constant; only V346\,Nor seems to fade. Our data on the
long-term evolution of V1057\,Cyg agree with the model predictions of
Kenyon \& Hartmann (1991) and Turner et al.~(1997) at near- and
mid-infrared wavelengths, but disagree at $\lambda\,{>}\,25\,\mu$m.
We discuss if this observational result at far-infrared wavelengths
could be understood in the framework of the existing models.
 }
          \maketitle


\section{Introduction} 
\label{sect:Intro} 

FU\,Orionis objects are low mass pre-main sequence stars defined as a
class by Herbig (1977). Most members of this class have undergone
outburst in optical light of 4 mag or more, followed by a fading phase
on the timescale of several decades. In some cases (e.g. Z\,CMa,
L1551~IRS\,5, BBW\,76) eruption in the optical was not observed but
they were identified as class members on the basis of their spectral
characteristics. For a comprehensive review of FU\,Orionis objects see
Hartmann \& Kenyon (1996).

According to the most widely accepted picture the FU\,Orionis outburst
is a consequence of a rapid temporal increase of the disk accretion
rate (Hartmann \& Kenyon 1996). Predictions of these types of models
(e.g. Kenyon \& Hartmann 1991, Bell et al. 1995, Turner et al. 1997)
have to be confronted with multiwavelength monitoring observations of
the outburst period and the fading phase. The fading phase is well
documented in the optical/near-infrared (Kopatskaya 1984, Ibragimov \&
Shevchenko 1988, Kenyon \& Hartmann 1991). Very few data have been
available so far at mid- and far-infrared wavelengths where thermal
emission of the disk and the circumstellar envelope can be observed.
Recently the {\it Infrared Space Observatory} (ISO, 1995$-$98, Kessler
et al.~1996) provided new photometric data in the $4.8{-}200\,\mu$m
range on FU\,Orionis-type stars.

In our study we search for systematic brightness variations during the
post-outburst phases of 7 FU\,Orionis objects at mid- and far-IR
wavelengths. We compiled two SEDs for each object: the first one is
based on IRAS photometry as well as ground-based observations and is
representative of the evolutionary status around 1983. The second SED
was compiled from observations taken around 1996$-$2000 including data
from ISOPHOT the infrared photometer on-board ISO (Lemke et al. 1996),
MSX (Egan \& Price 1996), and 2MASS (Cutri et al. 2003). The
comparison of the two SEDs provides information on the wavelength
dependence of the far-infrared flux evolution during a period of 15
years. As an example in the case of V1057\,Cyg we present a detailed
comparison of our results with the predictions of the models of Kenyon
\& Hartmann (1991) and Turner et al.~(1997).

\section{Target list, observations and data reduction} 

\begin{table*}
\begin{center}
\begin{tabular}{lcccccc}
\hline
Object                    &     Outburst       &    D [pc]               &    IRAS    &       MSX          &       ISO        &      2MASS       \\\hline
BBW\,76$^{1}$             & $<$1930$^{15}$     & 1700$^{18}$             & 07486-3258 & G248.7075-03.3686  &                  & 07503560-3306238 \\
CB34\,V$^{2}$             & $<$1994$^{2}$      & 1500$^{18}$             &            & G186.9520-03.8325  &                  & 05470377+2100347 \\
FU\,Ori$^{3}$             &  1937$^{15}$       & 450$^{18}$              & 05426+0903 &                    & SWS,LWS          & 05452235+0904123 \\
V1057\,Cyg$^{3}$          &  1970$^{15}$       & 600$^{18}$, 700$^{19}$  &            & G085.4595-01.0468  & PHT,LWS          & 20585371+4415283 \\
V1515\,Cyg$^{3}$          &  1950s$^{15}$      & 1000$^{18}$, 1050$^{3}$ & 20220+4202 & G079.9187+02.7391  & PHT              & 20234802+4212257 \\
L1551~IRS\,5$^{4}$       &    ?               & 140$^{18}$              & 04287+1801 &                    &                  & 04313407+1808049 \\
PP\,13S$^{5}$             &   $<$1900$^{16}$   & 300$^{18}$              & 04073+3800 &                    &                  & 04104088+3807517 \\
Parsamian\,21$^{6}$       &    ?               & 1800$^{18}$, 400$^{20}$ & 19266+0932 & G045.8149-03.8309  & PHT              & 19290085+0938429 \\
Re\,50\,N\,IRS1$^{7}$     &   1960-70$^{17}$   & 460$^{18}$              &            &                    & PHT,SWS          & ? \\
V1331\,Cyg$^{8}$          &    ?               & 550$^{18}$, 700$^{19}$  & 20595+5009 & G090.3121+02.6774  & LWS             & 21010920+5021445 \\
V1735\,Cyg$^{9}$          &  1957-65$^{15}$    & 900$^{18}$              & 21454+4718 & G093.7587-04.6371  & PHT,LWS,SWS      & 21472065+4732035 \\
V346\,Nor$^{10}$          & $\sim$1984$^{15}$  & 700$^{18}$              & 16289-4449 & G338.5458+02.1178  & CAM,PHT,LWS,SWS  & 16323219-4455306 \\
V883\,Ori$^{7}$           &    ?               & 460$^{18}$              & 05358-0704 &                   &                  & 05381810-0702259 \\
Z\,CMa$^{11}$             &    ?               & 930$^{18}$, 1150$^{21}$              & 07013-1128 & G224.6077-02.5574  & PHT,LWS,SWS      & 07034316-1133062 \\
RNO\,1B$^{12}$            &    ?               & 850$^{18}$              & 00338+6312 & G121.2940+00.6572  & CAM,LWS,SWS      & 00364599+6328529 \\
RNO\,1C$^{13}$            &    ?               & 850$^{18}$              & 00338+6312 & G121.2940+00.6572  & CAM,LWS,SWS      & 00364659+6328574 \\
AR\,6A$^{22}$             &    ?               & 800$^{22}$              &            & G203.2028+02.0653  &                  & 06405930+0935523 \\
AR\,6B$^{22}$             &    ?               & 800$^{22}$              &            & G203.2028+02.0653  &                  & 06405930+0935523 \\
OO\,Ser$^{14}$            &  1995$^{14}$       & 311$^{23}$              &            &                    & CAM,PHT,LWS,SWS  & 18294913+0116206 \\
V1647\,Ori$^{24}$         &  2004              & 400$^{25}$              & 05436-0007 &                    & CAM              & 05461313-0006048 \\
\hline
\end{tabular}
\caption{Up-to-date catalogue of FU\,Orionis objects (for column
  description see Sect.\,\ref{subsect:fuorcat}). References: [1]
  Eisl\"offel et al.~(1990); [2] candidate, other designation: V1184
  Tau, Yun et al.~(1997); [3] Herbig (1977); [4] Carr et al.~(1987);
  [5] Sandell \& Aspin (1998); [6] Staude \& Neckel (1992); [7]
  Reipurth 50 N IRS 1, Strom \& Strom (1993),
  $\alpha_{2000}=5^h40^m27.4^s$ $\delta_{2000}=-7^{\circ}27'33''$; [8]
  assumed to be in a pre-outburst state, McMuldroch et al.~(1992); [9]
  Elias (1978); [10] Graham \& Frogel (1985); [11] Hartmann et
  al.~(1989); [12] Staude \& Neckel (1991); [13] Kenyon et al.~(1993);
  [14] Hodapp (1995); [15] Hartmann \& Kenyon (1996); [16] Aspin \&
  Sandell (2001); [17] Reipurth \& Aspin (1997); [18] Sandell \&
  Weintraub (2001); [19] Chavarria-K. (1981); [20] Eiroa \& Hodapp
  (1990); [21] Herbst et al.~(1978); [22] candidate, Aspin \& Reipurth
  (2003); [23] de Lara et al.~(1991); [24] FU\,Ori candidate, in the
  state of rapid brightening since January 2004. An associated submm
  source was detected by Mitchell et al.~(2001); [25] Lis et
  al.~(1999) }
\label{tab:fuorcat}
\end{center}
\end{table*}
 
\subsection{Up-to-date catalogue of FU\,Orionis objects} 
\label{subsect:fuorcat}

When Herbig (1977) defined the class of FU\,Orionis objects, there
were only three known members. Since then a number of new objects have
been identified. As a first step of our study we compiled an
up-to-date list of FU\,Ori stars. We included FU\,Ori candidates
without attempting to homogenise the identification criteria of
different authors. The resulting catalogue, which is to a large extent
similar to the list of Sandell \& Weintraub (2001), is presented in
Tab.\,\ref{tab:fuorcat}. In this table Col.\,1 refers to the original
paper where the object was identified as FU\,Ori type. The next two
columns give the date of outburst (if known) and the distance to the
object, respectively. Associated sources in the IRAS and MSX
catalogues are given only in the case of good positional coincidence
with the optical coordinates. The RNO\,1B/1C system (separation
${\sim}5''$) was detected but unresolved by both IRAS and
MSX. Column\,6 shows which of the four ISO instruments (ISOCAM,
ISOPHOT, ISO-LWS, ISO-SWS) observed the source. These observations are
available for the public in the ISO Data
Archive\footnote{www.iso.vilspa.esa.es/ida/}. The last column
identifies the corresponding sources in the near-infrared 2MASS
survey. These source names were derived from the sexagesimal
coordinates thus give the equatorial positions for the epoch
2000. (For the position of Re\,50\,N\,IRS1 see figure caption.)

\subsection{Observations} 

Table\,\ref{tab:instruments} lists the sources of infrared photometric
data used in our study. The time distribution of the active periods of
the instruments/projects provides a possibility to check for long-term
variations of the infrared fluxes between $\sim$1983 (IRAS,
ground-based data) and 1996$-$2001 (ISO, MSX, 2MASS).

From the sources in Tab.\,\ref{tab:fuorcat} we selected those 7
objects for further study where enough data were available at both
epochs ($\sim$1983 and 1996$-$2000). Five of them have ISO multifilter
observations to create a complete mid/far-infrared SED. One object,
Parsamian\,21, was measured with ISOPHOT only at 65 and 100$\,\mu$m,
but MSX data are available at shorter wavelengths. V1331\,Cyg was not
observed by ISOPHOT, but the availability of MSX fluxes makes possible
the study of temporal variation at $\lambda \le 25\,\mu$m. The
detailed log of ISOPHOT observations is presented in
Tab.\,\ref{tab:ISOobs}.

\begin{table}
\begin{center}
\begin{tabular}{lllc}
\hline
Instrument   & Wavelengths [$\mu$m] & Aperture        &  Active period  \\ \hline
ground-based & JHKLMNQ              & $\leq 6''$      & $1970s{-}$     \\ 
IRAS         & $12, 25, 60, 100$      & $1{-}3'$         & $1983$         \\ 
MSX          & $4.25, 4.29, 8.28$     & $18''$          & $1996{-}1997$ \\ 
             & $12.13, 14.65, 21.34$  &                 &              \\ 
2MASS        & J, H, Ks             &                 & $1997{-}2001$ \\ 
ISOPHOT      & $4.8{-}120$             & $43''{-}180''$ & $1995{-}1998$ \\ \hline
\end{tabular}
\caption{Sources of infrared photometric data used in our study.}
\label{tab:instruments}
\end{center}
\end{table}

\begin{table*}
\begin{center}
\begin{tabular}{lccccl}
\hline
Object        &  Wavelengths [$\mu$m] &  Aperture [$''$]       &  Date         & Obs. mode &   ISO\_id          \\ \hline
V1057\,Cyg    &  $4.8, 12, 25, 60, 100$  & $180$         &  17-Apr-1996  & ON/OFF    &  15200607/15200608 \\
              &  $65, 100$               & $43{\times}43$   &   8-Nov-1996  & PHT32     &  35800611          \\ \hline
V1515\,Cyg    &  $4.8, 12, 25, 60$       & $180$         &  24-Apr-1996  & ON/OFF    &  15900605/15900606 \\
              &  $120$                   & $180{\times}180$ &  24-Apr-1996  & ON/OFF    &  15900613/15900614 \\ \hline
Parsamian\,21 &  $65, 100$               & $43{\times43}$   &  28-Sep-1996  & PHT32     &  31601103          \\ \hline
V1735\,Cyg    &  $4.8, 12, 25, 60, 100$  & $180$         &  21-Apr-1996  & ON/OFF    &  15600909/15600910 \\
              &  $4.8, 12, 25, 60, 100$  & $180$         &  20-May-1996  & ON/OFF    &  18501409/18501410 \\
              &  $65, 100$               & $43{\times43}$   &   3-Dec-1996  & PHT32     &  38301012          \\ \hline
V346\,Nor     &  $4.8, 12, 25, 60, 100$  & $180$         &   9-Feb-1996  & ON/OFF    &  08402303/08402304 \\ \hline
Z\,CMa        &  $4.8, 12, 25$           &  $52$         &   5-Nov-1997  & ON/OFF    &  72003006/72003005 \\
              &  $60, 100$               &  $99$         &   5-Nov-1997  & 5x1 scan  &  72003007          \\
\hline
\end{tabular}
\caption{Log of ISOPHOT observations. Observing modes are described in
Sect.~\ref{subsec:proc}, ISO\_id is the unique 8-digit identification
number of ISO observations.}
\label{tab:ISOobs}
\end{center}
\end{table*}

\subsection{ISOPHOT data processing} 
\label{subsec:proc}

Most observations were carried out as pairs of ON and OFF measurements
separated by $6{-}12'$. The $180''$ aperture was selected because it
was comparable to the IRAS far-infrared beam size. The far-infrared
flux of Z\,CMa was extracted from a $5{\times}1$ scan obtained with
the C100 camera. Three objects were measured at 65 and 100$\,\mu$m in
the high-resolution PHT32 mapping mode (see The ISO Handbook vol.~V,
Laureijs et al.~2001).

The data reduction was performed using the ISOPHOT Interactive
Analysis Software Package V10.0 (PIA, Gabriel et al. 1997). After
corrections for non-linearities of the integration ramps, the signals
were transformed to a standard reset interval. Then an orbital
dependent dark current was subtracted and cosmic ray hits were
removed. In case the signal did not fully stabilise during the
measurement time due to detector transients, only the last part of the
data stream was used. This was found mainly in observations with the
4.8$\,\mu$m and 12$\,\mu$m filters, while at other wavelengths the
measurements showed sufficient stability. The calibration of
measurements at 4.8, 12, and 25$\,\mu$m was performed by adopting the
default responsivity. At longer wavelengths first we tried comparison
with the on-board fine calibration source (FCS); however in most
ON/OFF observations the quality of the FCS measurement was
unacceptable and we returned to the default responsivity. As error
estimate we adopted an absolute calibration uncertainty of 25\%, which
represents well the sum of the random and systematic uncertainties.
Colour corrections were applied for each measurement by convolving the
observed SED with the ISOPHOT filter profile in an iterative way.

Three objects (V1057\,Cyg, V1735\,Cyg, Parsamian\,21) were measured at
65 and 100$\,\mu$m in the PHT32 observing mode (oversampled
map). Since this method is not scientifically validated we used our
own procedure for evaluation.  Instead of the highly interactive tool
implemented in PIA (Tuffs \& Gabriel 2003) we evaluated 23
observations of ISOPHOT standards (stars, asteroids, and planets) with
default processing and derived an empirical correction curve from the
comparison of the measured fluxes with their model predictions. The
fluxes were extracted from the maps using an algorithm like the one
developed for ISOPHOT mini-map observing mode (Mo\'or et
al. 2004). The typical uncertainty of the empirical correction does
not exceed the adopted 25\% error bars.

\subsection{IRAS data} 
\label{subsect:IRAS}

In order to have solid IRAS photometry for comparison with the ISOPHOT
data we have re-analysed the IRAS raw data of two FU\,Orionis objects
where the IRAS fluxes published by different authors (PSC, Weaver \&
Jones 1992, Kenyon \& Hartmann 1991) were discrepant. We utilised the
SCANPI processing tool at
IPAC\footnote{http://irsa.ipac.caltech.edu/applications/Scanpi/}. The
results are listed in Tab.\,\ref{tab:PHTflux}, remarks on individual
objects are given below.

\paragraph{V1057\,Cyg.}
This source was at the very edge of the IRAS coverage and only few
scans are available. At 12$-$60$\,\mu$m we adopted the fluxes
published by Weaver \& Jones (1992), though their formal uncertainties
seem to be too optimistic. At 100$\,\mu$m, where only two detectors
crossed the star, the source is surrounded by structures which are
equally strong or stronger than V1057\,Cyg itself. The flux
uncertainty is fully dominated by the accuracy of background
determination. Experiencing with different background subtractions we
estimate $47{\pm}10$\,Jy for V1057\,Cyg's flux density at 100$\,\mu$m.

\paragraph{V1515\,Cyg.}
In this case the background is even more structured than around
V1057\,Cyg, and the coarse spatial resolution of IRAS is a problem
already at 12$\,\mu$m. At 12 and 25$\,\mu$m we estimate
$3.7{\pm}1.0$\,Jy and $6.8{\pm}2.1$\,Jy, respectively, as integrated
flux densities which also include an extended structure around the
source, probably the infrared equivalent of the ring one can see in
the optical wavelengths. The formal error bars reflect the uncertainty
related to the background subtraction. At 60$\,\mu$m the lack of
spatial resolution is the limiting factor. The emission peak is offset
by about 20$''$ from the nominal position of V1515\,Cyg, indicating
that part of the emission from the region is unrelated to the
star. For flux density towards the position of V1515\,Cyg our estimate
is $25{\pm}10$\,Jy. At 100$\,\mu$m the emission is dominated by a
local peak at about 1$'$ from the position of V1515\,Cyg, and only an
upper limit of $110$\,Jy can be derived for the FU\,Orionis object.


\section{Results} 

\begin{figure*}  
\begin{center} 
\psfig{figure=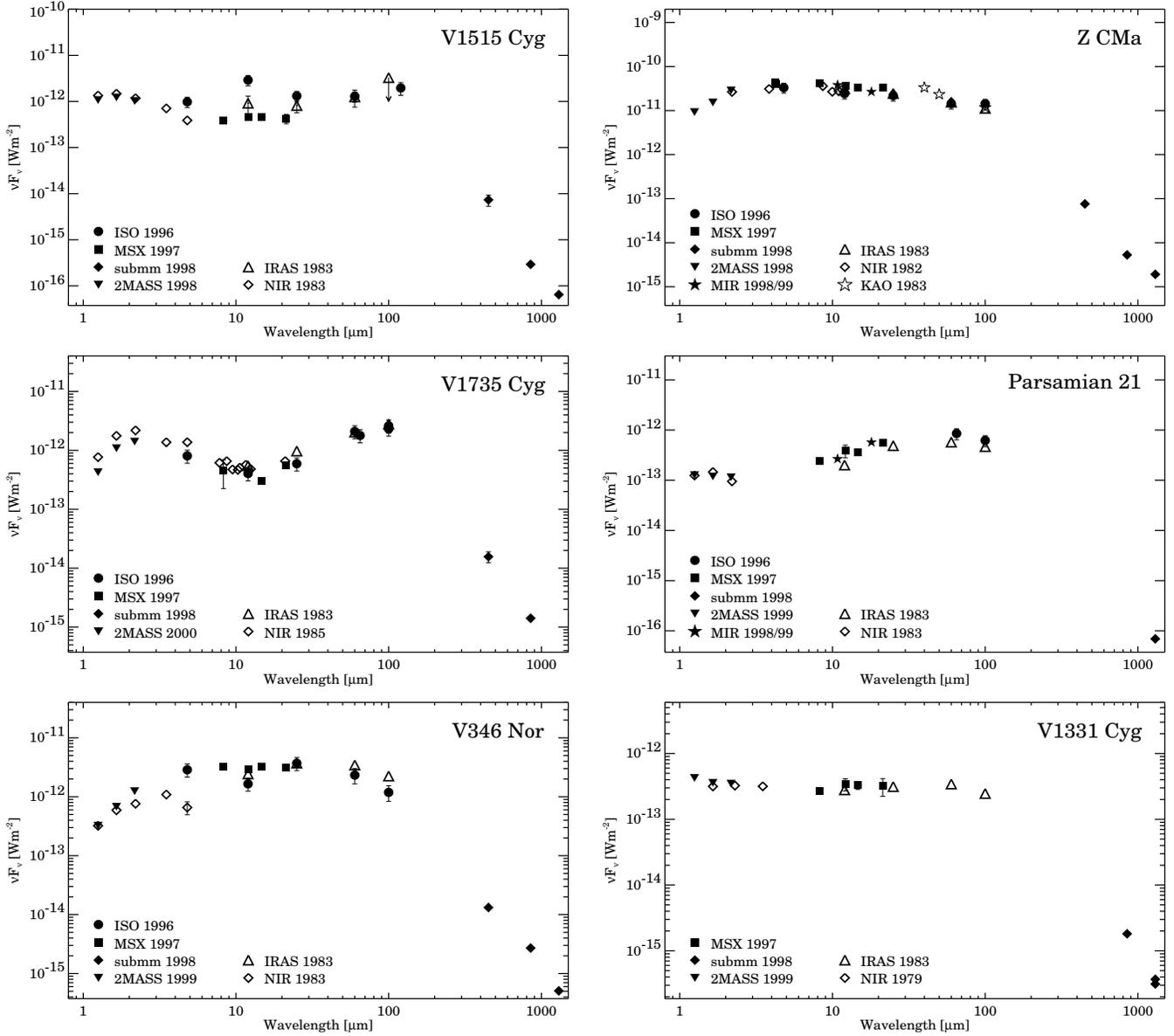,width=180mm,height=160mm}
\caption{Spectral energy distributions of 6 FU\,Orionis-type
objects. Open symbols correspond to an epoch of $\sim$1983 while
filled symbols mark the SED as of 1996$-$2000. The data are presented
with no reddening correction. Error bars smaller than the symbol size
were not plotted. For V1331\,Cyg there are no NIR data from 1983; we
plotted data from 1979 (the K-band lightcurve shows constant flux in
this period). In the case of V346\,Nor the ISO and IRAS beams also
include the nearby pre-main sequence star Re\,13. For comparison the
plotted MSX flux densities are sums of the individual fluxes of the
two objects.}
\label{fig:main}
\end{center} 
\end{figure*}

\begin{figure}  
\begin{flushright} 
\psfig{figure=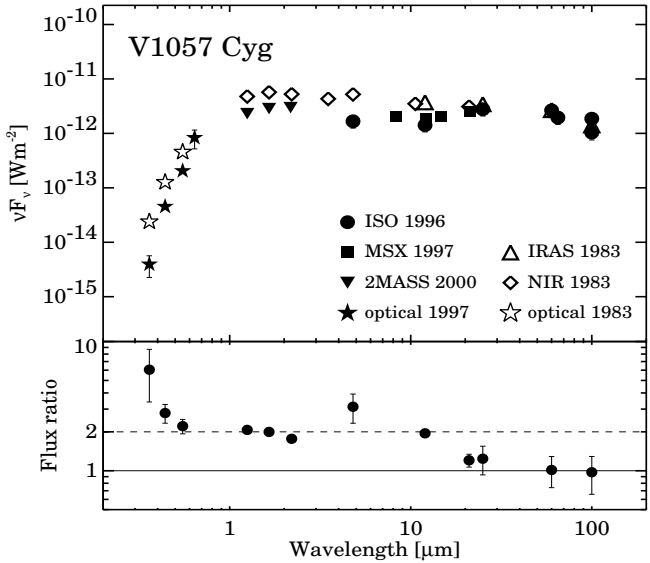,width=86mm,height=75mm}
\caption{Spectral energy distribution of V1057\,Cyg. Open and filled
symbols are as in Fig.~\ref{fig:main}. No reddening correction was
performed. The lower panel shows the ratios of fluxes obtained at the
two epochs ($\sim$1983 to 1996$-$2000).}
\label{fig:V1057Cyg}
\end{flushright} 
\end{figure}

The results of the ISOPHOT photometry as well as our newly derived IRAS
flux densities are presented in Tab.\,\ref{tab:PHTflux}.
Figure\,\ref{fig:main} displays the SEDs of the sources from the two
different epochs ($\sim$1983 and 1996$-$2000). The results for V1057\,Cyg,
supplemented with optical data, are shown in Fig.\,\ref{fig:V1057Cyg}.

By comparing the two SEDs of each source one can see the following
evolutionary trends:
\begin{itemize}
\item at near-IR wavelengths ($\lambda\,{\leq}\,5\,\mu$m) the sources
  show various behaviour. Parsamian\,21, V1331\,Cyg and Z\,CMa are
  unchanged, V1057\,Cyg, V1515\,Cyg and V1735\,Cyg have faded,
  V346\,Nor has become slightly brighter.
\item in the mid-IR ($5\,{\leq}\,\lambda\,{\leq}\,20\,\mu$m) only
  V1057\,Cyg shows systematic flux change: it became fainter by a
  factor of 2 during the period (lower panel in
  Fig.\,\ref{fig:V1057Cyg}).
\item at far-IR wavelengths ($\lambda\,{\geq}\,60\,\mu$m) five stars
  (V1057\,Cyg, V1735\,Cyg, Z\,CMa, Par\,21, and V1515\,Cyg) remained
  constant while V346\,Nor seems to have become fainter. For
  V1331\,Cyg there are no FIR data other than IRAS.
\end{itemize}
Below we make remarks on the individual sources. The case of
V1057\,Cyg will be discussed in details in
Sect.\,\ref{sect:Discussion}.

\paragraph{V1515\,Cyg.}

\begin{figure}  
\begin{flushright} 
\psfig{figure=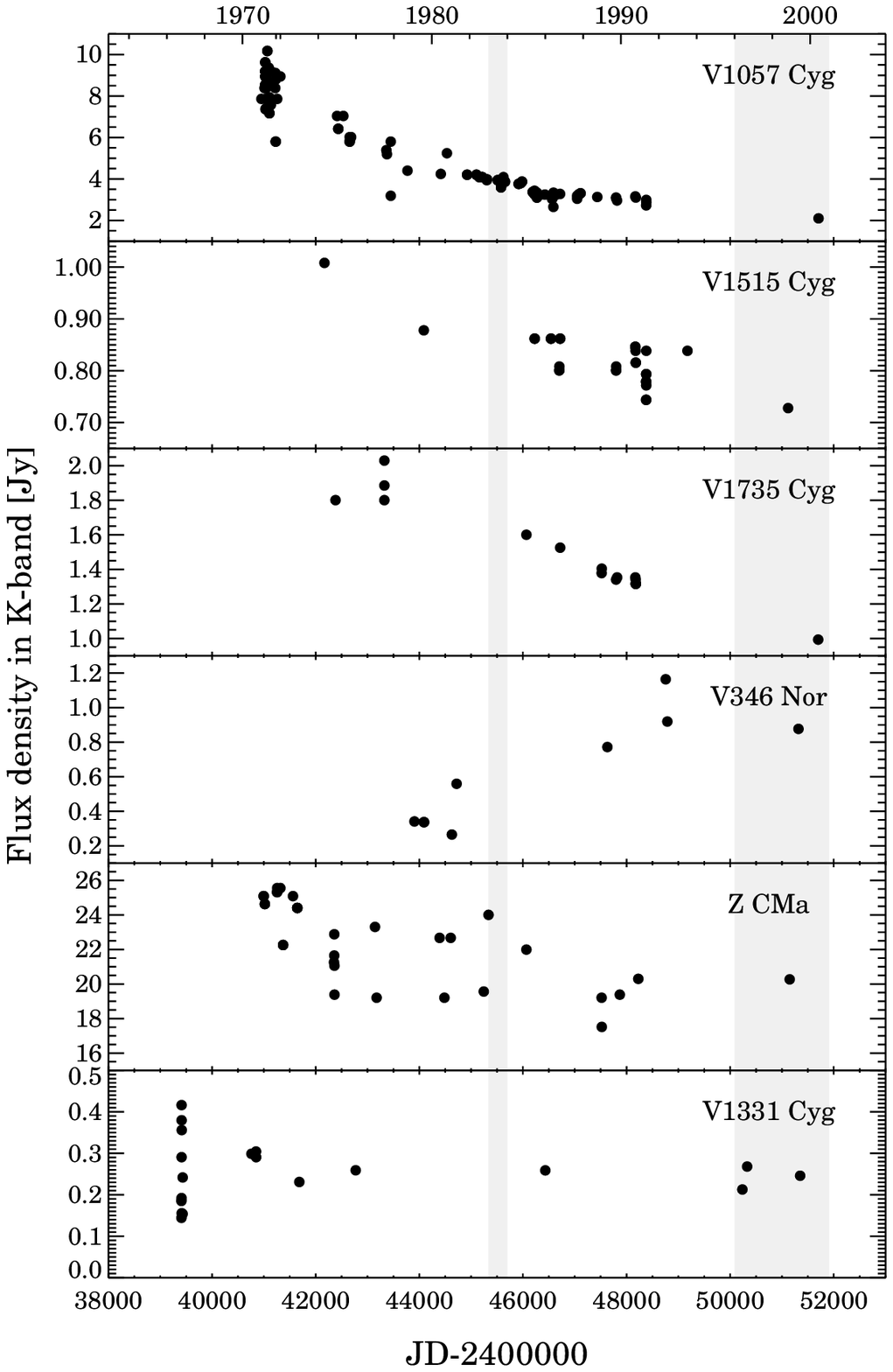,width=86mm,height=130mm,angle=0}
\caption{K-band lightcurve of 6 FU\,Orionis objects for the period
1963$-$2003. Data are taken from the literature (for Parsamian\,21 no
observations are available). Grey stripes mark the two epochs 
(1983 and 1996$-$2000) which form the baseline for the long-term
evolution.}
\end{flushright} 
\label{fig:klight}
\end{figure}

According to the 850$\,\mu$m submillimetre map of Sandell \& Weintraub
(2001) this source is situated in the middle of a cold arcminute-size
core, and this extended emission is also visible on the MSX
8.28$\,\mu$m map. In addition the 2MASS and MSX data reveal a compact
object 2MASS 20235198+4211260 at about $80''$ from V1515\,Cyg. Since
this source is not included in the MSX Point Source Catalog, we
extracted fluxes from the MSX maps using aperture photometry. The
resulting SED peaks at ${\approx}4\,\mu$m and its position on the
J$-$H vs.~H$-$K diagram suggests a reddened background star with
$A_V\,{\geq}\,15$ mag. The contribution of this source in the $3'$
ISOPHOT beam can explain why the ISOPHOT 4.8$\,\mu$m point is higher
than the M-band observation of 1983. In the 12$-$25$\,\mu$m range ISO
fluxes are higher while the MSX fluxes are lower than IRAS. This
result, however, can be explained by the presence of the arcminute
core associated to V1515\,Cyg, taking into account the different beam
sizes of the three instruments (MSX: $18''$, IRAS: $1'$, ISOPHOT:
$3'$). At near-infrared wavelengths the data indicate a slight
($\sim$~15\%) flux decrease, which is also visible in the K-band
lightcurve (Fig.\,\ref{fig:klight}). In the mid-infrared the
comparison of the 1983 and 1996$-$98 SEDs are complicated by the beam
confusion. The fact, however, that at ${\sim}20\,\mu$m Q-band
measurements from 1974 (4.1$\pm$0.7\,Jy, Cohen 1974) and 1989
(2.9$\pm$0.6\,Jy, Kenyon et al. 1991) agree with the MSX flux value of
3.0$\pm$0.7\,Jy, suggests that the mid-infrared fluxes of V1515\,Cyg
remained constant. At far-infrared wavelengths the emission of
V1515\,Cyg was constant within the measurement uncertainties.

\paragraph{V1735\,Cyg.}

The 850$\,\mu$m submillimetre map of Sandell \& Weintraub (2001)
reveals a cold compact source, V1735\,Cyg SM\,1 at $20''$ from the
star dominating the emission at both 450 and 850$\,\mu$m. From a
careful analysis of the MSX images we concluded that in the whole
near- and mid-infrared regime the submillimetre source is invisible
and the observed emission can be assigned to V1735\,Cyg. Thus, the
star is responsible for the observed flux decrease of
${\approx}\,40$\% below 5$\,\mu$m, in accordance with the change in
the K-band lightcurve (Fig.\,\ref{fig:klight}). No flux variation is
visible at mid-infrared wavelengths. The relative contribution of the
submm source, however, is probably increasing at longer wavelengths,
as indicated by the fact that the IRAS position is located between the
positions of the FU\,Orionis object and that of SM\,1. Both the IRAS
and ISOPHOT beams include both sources. Our result of no flux
variation at far-infrared wavelengths indicates that the emission of
V1735\,Cyg was constant.

\paragraph{V346\,Nor.}

This star is situated in a complex region including the PMS star
Reipurth\,13, HH\,56/57 and some cold extended emission within a radius
of $60''$ (Sandell \& Weintraub, 2001). Applying a maximum entropy
based deconvolution technique Prusti et al. (1993) were able to
determine the relative contribution of Re\,13 (separated by
${\approx}\,50''$ from the FU\,Orionis star) in the IRAS bands, and
found that its flux is a factor of 3-5 times lower than the emission
of V346\,Nor. At near-infrared wavelengths the FU\,Orionis star -- in a
unique way in our sample -- became brighter. This behaviour, reported
already by Prusti et al.~(1993), is also shown by the K-band
lightcurve (Fig.\,\ref{fig:klight}). The brightening seems to be
wavelength dependent: its amplitude increases monotonically from J to
K. The same trend can still be observed at mid-infrared wavelengths up
to 25$\,\mu$m where the IRAS and ISOPHOT fluxes are the same. On the
other hand at $\lambda\,{\geq}\,60\,\mu$m the observed fluxes became
lower between 1983 and 1996. Assuming that Re\,13 (which is also
included in the beam) is not variable we may conclude that V346\,Nor
faded at 60 and 100$\,\mu$m. It is remarkable that the flux change is
associated with a change in the far-infrared colour, too, which became
warmer during the 15 years. The observed change could be an
indication for a temperature rise in the outer parts of the
circumstellar environment, but it might also be an artifact caused by
the different sensitivities of the IRAS and ISOPHOT instruments for
the extended emission.

\paragraph{Z\,CMa} is a binary system with a separation of 0.1$''$. The 
component which dominates at infrared wavelengths is probably a Herbig
Ae/Be star, while its optically brighter companion is an FU\,Orionis
object. The system is embedded in an elongated disk-like submm core
(Sandell \& Weintraub 2001), and was unresolved for both IRAS and
ISOPHOT. According to the K-band lightcurve (Fig.\,\ref{fig:klight})
Z\,CMa did not change after 1974. The same conclusion can be drawn
for the whole 2$-$100$\,\mu$m SED. This result, however, does not
exclude small flux variation of the FU\,Orionis component since the
infrared SED of the system is dominated by the intermediate mass
component.

\paragraph{Parsamian\,21.} This source is relatively isolated, no extended 
emission is associated with it at submillimetre wavelengths (Henning
et al. 1998), thus beam differences of the different instruments are
unimportant. The comparison of the two SEDs does not reveal any long
term flux variation between 1983 and 1996$-$98 within the measurement
uncertainties.

\paragraph{V1331\,Cyg} is believed to be in a state prior to an 
FU\,Orionis-type outburst (Welin 1976, McMuldroch et al. 1993). The
source is compact at submillimetre wavelengths (Henning et al. 1998),
thus the comparison of the various observations is not influenced by
beam effects. Our data show no evidence of temporal variation of the
source's flux in the whole infrared regime. The K-band lightcurve
(Fig.\,\ref{fig:klight}) is consistent with this conclusion.

\begin{table*}
\begin{center}
\begin{tabular}{cccccccc}
\hline
$\lambda$ [$\mu$m] & V1057\,Cyg          & V1515\,Cyg        & V1735\,Cyg         & V346\,Nor         & Z\,CMa              & Parsamian\,21 & V1331\,Cyg \\ \hline
$4.8$                & $2.67 \pm 0.67$    & $1.57 \pm 0.39$   & $1.29 \pm 0.32$    & $ 4.59 \pm 1.15$  & $53.16 \pm 13.29$   & &  \\
$12$                 & $5.68 \pm 1.42$    & $11.64 \pm 2.91$  & $1.62 \pm 0.40$    & $6.61 \pm 1.65$   & $97.33 \pm 24.33$   & &  \\
$25$                 & $23.19 \pm 5.80$   & $11.01 \pm 2.75$  & $4.94 \pm 1.24$    & $30.82 \pm 7.71$  & $182.8 \pm 45.7$  & &   \\
$60$                 & $52.99 \pm 14.11$  & $25.80 \pm 6.45$  & $41.76 \pm 10.44$  & $46.54 \pm 13.45$ & $290.1 \pm 72.5$   & &   \\
$65^*$               & $42.26 \pm 10.57$  &                   & $38.68 \pm 9.67$   &                   &                     & $18.51 \pm 4.63$ & \\
$100$                & $34.50 \pm 9.28$   &                   & $77.23 \pm 19.31$  & $39.52 \pm 11.65$ & $479.1 \pm 119.8$ & &  \\
$100^*$              & $62.06 \pm 15.52$  &                   & $86.82 \pm 21.71$  &                   &                     & $20.71 \pm 5.18$ &  \\
$120$                &                    & $78.45 \pm 24.03$ &                    &                   &                     & &    \\
\hline
IRAS-12              & $14.89 \pm 0.11$   & $3.7 \pm 1.0$   & $2.19 \pm 0.20$    & $9.73 \pm 0.39$   & $125.1 \pm 5.0$   & $0.80 \pm 0.06$  & $1.12 \pm 0.03$ \\
IRAS-25              & $28.72 \pm 0.08$   & $6.8 \pm 2.1$   & $8.09 \pm 0.52$    & $30.98 \pm 1.33$  & $204.5 \pm 8.7$   & $4.07 \pm  0.26$ & $2.62 \pm 0.02$ \\
IRAS-60              & $53.71 \pm 2.48$   & $25 \pm 10$ & $40.84 \pm 4.90$   & $69.07 \pm 4.83$  & $312.3 \pm 40.6$  & $11.49 \pm 1.03$ & $6.88 \pm 0.23$ \\
IRAS-100             & $47.0 \pm 10.0$    & $< 110$      & $92.97 \pm 15.63$  & $74.91 \pm 4.20$  & $375.5 \pm 63.2$  & $15.58 \pm 2.04$ & $8.22 \pm 1.19$ \\ \hline
\end{tabular}
\caption{Photometric results of ISOPHOT and IRAS. All fluxes are
colour-corrected and presented in Jy units. (*): PHT32 observing mode.
Source of IRAS data: PSC -- V1735\,Cyg, V346\,Nor, Z\,CMa,
Parsamian\,21; Weaver \& Jones (1992) -- V1331\,Cyg; this work
(Sect.\,\ref{subsect:IRAS}) -- V1515\,Cyg.  In the case of V1057\,Cyg
the $12{-}60\,\mu$m fluxes are from Weaver \& Jones (1992), the flux
at $100\,\mu$m was re-determined in this work
(Sect.\,\ref{subsect:IRAS}).}
\label{tab:PHTflux}
\end{center}
\end{table*}


\section{Discussion: the case of V1057\,Cyg}
\label{sect:Discussion}

In this section we compare our new observational results on the
temporal evolution of FU\,Orionis objects with predictions of models
developed to describe their circumstellar structure. From our sample
V1057\,Cyg has the best documented multiwavelength flux evolution
following its outburst in 1970. The ISO data points, supplemented
with other infrared observations (Fig.\,\ref{fig:V1057Cyg}), reveal a
fast temporal evolution between $\sim$1983 and 1996$-$2000 (the
fastest rate in the sample). The wavelength dependence of the
evolution is also well determined. Since detailed models of
V1057\,Cyg are available in the literature, we chose this object to
compare the new observational results on the temporal evolution with
model predictions.

As far as we know there are two models of the circumstellar
environment of V1057\,Cyg to fit the complete 1$-$100\,$\mu$m infrared
SED (Kenyon \& Hartmann 1991, Turner et al.~1997). Kenyon \& Hartmann
(1991) collected and analysed all data available until 1991 and
modelled the post-outburst evolution in the framework of an accretion
disk model. They concluded that a simple accretion disk model cannot
explain the large far-IR fluxes, and investigated two possible
geometries, a flared disk and an extended infalling envelope. Since
the flared disk would require very high coverage factor, the authors
favour a model where a flat disk is embedded in a spherically
symmetric envelope with a wind-driven polar hole. The envelope, which
is the remnant of the molecular cloud core, reprocesses the radiation
from regions close to the central star. Turner et al. (1997) fitted
the SED of V1057\,Cyg by computing outbursting flared disk models in
which the mass flux varies with radius. The model includes
reprocessing of disk emission by other parts of the disk. An envelope
with a central hole exposing the inner disk is included as a layer of
uniform thickness on the top of the disk.

At optical and near-IR wavelengths ($\lambda\,{\leq}\,2.2\,\mu$m) both
models claim that emission of the central source (the star plus the
innermost part of the accretion disk) dominates the observed flux.
After the outburst the accretion rate close to the star decreases
leading to a flux decrease in this wavelength range. Our data confirm
that the flux decay was observable between 1983 and 2000, too. The
data reveal that in the R, J, H, and K bands the flux dropped by the
same constant factor of 2, while at B and V a larger decay was
observed (lower panel of Fig.\,\ref{fig:V1057Cyg}). This wavelength
dependence may reflect the drop of effective temperature leading to a
shift in the peak of the emission of the central source towards longer
wavelengths.

Between 3 and 10$\,\mu$m the origin of the emission is the release of
accretion energy in the disk (Kenyon \& Hartmann 1991), and also
starlight reprocessed in the same part of the disk and in an envelope
(Turner et al.~1997). Decreasing accretion rate at the center (which
leads to a decrease in the bolometric luminosity) would cause the drop
of emission of all three components. The timescale at which viscous
accretion could change is related to the dynamical timescale at this
distance (several years at $\sim$1 AU, Pringle 1981). The timescale
on which the reprocessed emission follows the fading of the central
source depends on the optical depth of the reprocessing medium. For
optically thin medium (disk surface layer, envelope) an almost
instantaneous reaction is expected (several hours, Chiang \& Goldreich
1997, Eq.\,25). For optically thick emission (disk interior) the
thermal timescale related to the gas component dominates (several
years at $\sim$1 AU for a typical T Tauri star, Chiang \& Goldreich
1997, Eq.\,28). In the case of V1057\,Cyg the 3$-$10\,$\mu$m emission
is probably not completely optically thick, as indicated by the
presence of weak spectral features (silicate emission at 9.7$\,\mu$m,
Wooden et al.~1995, Hanner et al.~1998). Since our temporal baseline
(1983$-$1998) significantly exceeds the mentioned timescales and since
the post-outburst evolution started already in the seventies, the
model prediction is that the 3$-$10\,$\mu$m emission is decreasing
synchronised with the rate of the optical/near-IR decay in a
wavelength independent way. Our measurements are fully consistent
with these predictions. (Fig.\,\ref{fig:V1057Cyg} lower panel)

The emission at $\lambda\,{>}\,10\,\mu$m is reprocessed starlight in
both the model of Kenyon \& Hartmann (1991) and of Turner et
al.~(1997). The infrared radiation emerges from an envelope where
dust particles are in radiative equilibrium with the illumination from
the central region (in the model of Turner et al. the envelope is
optically thick for its own radiation at $\lambda\,{\le}\,30\,\mu$m).
The temporal evolution implied by the models is similar to the
predicted trend at shorter wavelengths: a wavelength independent
fading at the same rate as the fading of the central illuminating
source (although the time-lags are gradually longer at longer
wavelengths, Chiang \& Goldreich 1997). Our results are not fully
consistent with this picture. Though at 10\,$\mu$m the flux dropped
between 1983 and 1996 by a factor of 2 similarly to the
optical/near-IR rate, at far-IR wavelengths
($\lambda\,{\ge}\,60\,\mu$m) the flux of V1057\,Cyg remained
constant. The comparison of the IRAS 60 and 100$\,\mu$m photometric
points with our new ISOPHOT data at the same wavelengths clearly
demonstrates that the far-infrared fluxes of V1057\,Cyg showed no
variation between 1983 and 1997. The transition between the two types
of behaviour is around 25\,$\mu$m.

Our results indicate that -- unlike in the above mentioned models -- at
$\lambda\,{\ge}\,10\,\mu$m two important emission components have to be
taken into consideration. Between 10 and 25\,$\mu$m we probably
observe the envelope, while at longer wavelengths the emission cannot
be explained this way. The material responsible for the
$\lambda\,{>}\,25\,\mu$m emission is apparently {\it not} an 
optically thin medium reprocessing the light of the central source,
because its radiation does not react immediately on the source's fading. 
Moreover, the emitting material has the following
properties: (1) it is relatively cold; (2) it has a flat spectrum below 
100\,$\mu$m, suggesting a $T\,{\sim}\,r^{-0.5}$ radial temperature profile.

The first possibility for the nature of the far-infrared emitting
component could be a reprocessing medium
which is optically thick (e.g. a flared disk), and its time-lag behind
the fading of the central source significantly exceeds our temporal
baseline. Kenyon \& Hartmann (1991), however, demonstrated that the
outer part of a flared disk around V1057\,Cyg would produce an unacceptably 
large covering factor and rejected this possibility. 
In addition, based on more detailed calculations, Bell et al. (1997) 
pointed out that such a flared structure would not be realised in nature. 
At a radius of $\sim$\,10\,AU the flared disk would develop
into a 'decreasing curvature' disk, corresponding to the condensation
of water ice, thus the outer disk is shadowed from the illumination of 
the central source.
Such a disk cannot produce either the $T\,{\sim}\,r^{-0.5}$ 
temperature profile or the high level of far-infrared emission.
Based on these arguments we exclude reprocessing of the radiation of the central
source in the outer part of the disk as the origin of the far-infrared emission.

Nevertheless, it may be interesting to mention that the critical radius 
where the disk turns 
to concave (i.e. the border line between the illuminated and shadowed parts 
of the disk) corresponds to a temperature of $\approx$300\,K 
(Bell et al. 1997). 
Since the blackbody emission of this temperature contributes mainly
to the 25\,$\mu$m range of the SED, one might speculate that the change in the
temporal behaviour of the SED of V1057\,Cyg, also at 25\,$\mu$m, could be related
to the turnover of the disk profile. However, in the previous paragraphs we 
concluded that neither the short wavelength emission
(${\lambda}{\leq}25\,{\mu}$m, attributed an envelope) nor the long wavelength 
part of the SED are dominated by the radiation of an optically thick passive 
disk. We think that the agreement of the transition wavelength in  
V1057\,Cyg at 25\,$\mu$m with the peak of the blackbody curve at the  
sublimation temperature of water ice in the disk is a random coincidence.  
This conclusion is further supported by the case of V1735\,Cyg, where 
the shortest wavelength of constant flux is at 10\,$\mu$m, clearly different from
the temperature of the turnover radius (Fig.\,\ref{fig:main}).

As a second possibility for the origin of the far-infrared emission
Bell et al. (1997) proposed that the heating of the outer part of an optically 
thick disk could be
dominated by the ambient radiation field of the cloud core around the young
star, rather than by the radiation of the central source. As a result, 
the radial
temperature profile becomes constant at the ambient temperature value,
producing a shallower -- or even approximately flat -- far-infrared SED. 
Since its origin is unrelated to the central source, the 
far-infrared emission is not expected to vary in time.

A third possibility would be if the far-IR emission is powered by accretion 
in the outer disk ({\it active disk}), where the accretion
rate is constant and unrelated to the outburst of the central region
(Bell et al. 1995). However, the spectral shape of such a standard accretion
disk would follow the canonical
${\nu}F_{\nu}\,{\sim}\,{\lambda}^{-4/3}$ law, in contradiction with
the observed flat spectrum (Fig.\,\ref{fig:V1057Cyg}). 
A possible solution for this problem could be if the temperature profile 
of the accretion 
disk differs from the standard one. Lodato \& Bertin (2001) suggested that
in a self-gravitating protostellar disk the non-Keplerian rotation
curve may result in a nonstandard temperature distribution, and thus
in a flat spectrum in the far-IR. The high disk mass required by this
model (${\approx}\,1\,M_{\sun}$) is not inconsistent with measurements of
some FU\,Orionis objects (Henning et al.~1998, 
Sandell \& Weintraub 2001), but the measured
disk mass of V1057\,Cyg (${\approx}\,0.1\,M_{\sun}$, Sandell \& Weintraub 2001)
seems to be too low causing any observable effect.

The last possibility would be to assume that the origin of the far-IR emission
is unrelated to V1057\,Cyg. Such a source could be a -- so far undetected --
embedded IR companion. It is interesting to note that Bell et al. (1995) already
speculated about such a companion which could have triggered the outburst of
V1057\,Cyg by a close passage.  A deeply embedded  infrared companion,
like e.g. the very red object discovered in the vicinity of  LkH$\alpha$\,198
at 10$\mu$m (Lagage et al. 1993), might be too faint to be
detected at near-infrared wavelengths so far. At mid- and far-infrared
wavelengths  V1057\,Cyg  has never been imaged with high-resolution
instruments, thus the existence of a companion unresolved in the beam of the
MSX (18$''$) cannot be excluded.

In the previous paragraphs we argued that an optically thick disk, reprocessing 
the radiation of the central source, or accretion with a non-Keplerian 
rotation curve are less likely explanations, but a shallower temperature profile
dictated by the ambient radiation field or the possibility of an unrelated
infrared companion cannot be excluded. The observations (Fig.\,1) suggest that
the constancy of far-infrared emission might be a general property of
FU Orionis objects. Though a growing number of FU Orionis 
objects are found to be binaries (Reipurth \& Aspin 2004), it is not proved 
yet that {\it all} FU Orionis objects have companions. Thus we think that 
the most likely origin of the far-infrared radiation in V1057\,Cyg (and perhaps
in other FU Orionis objects) is emission
from the outer part of an optically thick disk whose temperature profile is
controlled by the ambient radiation field.

Our investigation demonstrates that temporal variability of the
infrared emission can provide important information on the physical
properties of circumstellar material around FU\,Orionis objects. This
kind of diagnostics of the disk structure -- proposed first by Chiang
\& Goldreich (1997) -- can also be applied on any other types of young
stellar objects where the energy source of the disk, either accretion
or reprocessing, is time-variable (e.g.~exors). New infrared data on
some FU\,Orionis objects are expected from the Spitzer Space Telescope
which could show whether the evolutionary trends outlined by the ISO
data also continue in our days, and whether the variation of the SEDs
at longer wavelengths are responding to the flux drop at shorter
wavelengths.


\begin{acknowledgements}
  The observations were reduced using the ISOPHOT Interactive Analysis
  package PIA, which is a joint development by the ESA Astrophysics
  Division and the ISOPHOT Consortium, lead by the Max-Planck-Institut
  f\"ur Astronomie (MPIA). This research has made use of the NASA/
  IPAC Infrared Science Archive, which is operated by the Jet
  Propulsion Laboratory, California Institute of Technology, under
  contract with the National Aeronautics and Space Administration.
  This research has made use of the SIMBAD database, operated at CDS,
  Strasbourg, France. We received comments and suggestions from
  L.~Szabados and D.~Apai who carefully read the manuscript. 
  We are grateful to an anonymous referee who called our attention to
  models of the temperature and opacity structure in the outer parts of
  protostellar disks. The work
  was partly supported by the grant OTKA T\,037508 of the Hungarian
  Scientific Research Fund. P.\'A. acknowledges the support of the
  Bolyai Fellowship.
\end{acknowledgements} 



\end{document}